\documentclass[a4paper,fleqn,usenatbib,twocolumn]{mnras}

\usepackage{graphicx}	
\usepackage{amsmath}	
\usepackage{amssymb}	
\usepackage{savesym}
\usepackage{natbib}

\usepackage{mathrsfs,dsfont}
\usepackage{multirow}
\usepackage{float}
\usepackage{booktabs}
\usepackage{epstopdf}
\usepackage{color}
\usepackage{hyperref}
\usepackage{verbatim}
\usepackage{soul}
\usepackage[none]{hyphenat}

\graphicspath{{figures/}}

\def\be{\begin{equation}}
\def\ee{\end{equation}}
\def\kms{{\rm \,km\,s^{-1}}}

\def\Myr{{\rm \,Myr}}
\def\Gyr{{\rm \,Gyr}}

\def\Mpc{{\rm \,Mpc}}
\def\kpc{{\rm \,kpc}}

\def\keV{{\rm \,keV}}

\def\msun{{\,M_\odot}}

\newcommand{\dd}{{\rm d}}

\usepackage{xcolor}

\title[N-waves in Merging Clusters]{Pairs of Giant Shock Waves ({\em N}-Waves) in Merging Galaxy Clusters}

\author[Congyao Zhang et al.]{
Congyao Zhang,$^{1}$\thanks{E-mail: cyzhang@astro.uchicago.edu}
Eugene Churazov,$^{2,3}$
Irina Zhuravleva$^{1}$
\\
$^1$~Department of Astronomy and Astrophysics, University of Chicago, Chicago, IL 60637, USA \\
$^2$~Max Planck Institute for Astrophysics, Karl-Schwarzschild-Str. 1, D-85741 Garching, Germany  \\
$^3$~Space Research Institute (IKI), Profsoyuznaya 84/32, Moscow 117997, Russia \\
}

\date{Accepted XXX. Received YYY; in original form ZZZ}

\pubyear{2019}

\begin{document}
\label{firstpage}
\pagerange{\pageref{firstpage}--\pageref{lastpage}}
\maketitle

\begin{abstract}
When a subcluster merges with a larger galaxy cluster, a bow shock is driven ahead of the subcluster. At a later merger stage, this bow shock separates from the subcluster, becoming a ``runaway'' shock that propagates down the steep density gradient through the cluster outskirts and approximately maintains its strength and the Mach number. Such shocks are plausible candidates for producing radio relics in the periphery of clusters. We argue that, during the same merger stage, a secondary shock is formed much closer to the main cluster center. A close analog of this structure is known in the usual hydrodynamics as {\em N}-waves, where the trailing part of the ``{\em N}'' is the result of the non-linear evolution of a shock. In merging clusters, spherical geometry and stratification could further promote its development. Both the primary and the secondary shocks are the natural outcome of a single merger event and often both components of the pair should be present. However, in the radio band, the leading shock could be more prominent, while the trailing shock might conversely be more easily seen in X-rays. The latter argument implies that for some of the (trailing) shocks found in X-ray data, it might be  difficult to identify their ``partner'' leading shocks or the merging subclusters, which are farther away from the cluster center. We argue that the Coma cluster and A2744 could be two examples in a post-merger state with such well-separated shock pairs.
\end{abstract}



\section{Introduction} \label{sec:introduction}

Intracluster medium (ICM) is the hot gaseous component of galaxy clusters, which is composed of fully ionized, magnetized plasma. While magnetic fields play an important role in the ICM, pure fluid dynamics of an ideal gas provides a reasonable first-order approximation of its properties on macro scales. Many textbook hydrodynamic phenomena associated with gas motions (e.g. shocks, cold fronts, turbulence) have been identified in galaxy clusters through their X-ray and Sunyaev-Zeldovich (SZ) signals \citep[e.g.][]{Markevitch2002,Ghizzardi2010,Mascolo2019}.
These motions and the ICM distribution itself are tightly connected with the growth history of galaxy clusters and provide an opportunity to explore the cluster formation theory (see e.g. \citealt{Kravtsov2012,Vikhlinin2014,Simionescu2019} for reviews).

The ICM is largely heated via the thermalization of the gravitational energy of the infalling matter. Therefore, the virial velocity of a cluster is comparable to the ICM sound speed and can be much larger than the sound speed of the intergalactic medium falling into the cluster for the first time. Shock waves are thus arising naturally during the merger processes \citep[see reviews by][]{Markevitch2007,Vikhlinin2014}. Focusing mainly on minor mergers, \citet{Zhang2019a,Zhang2019b} have suggested distinguishing two phases in the formation and evolution of merger shocks in galaxy clusters, specializing for the case of a minor merger. In the \emph{driven} phase, a bow shock is formed ahead of a smaller infalling subcluster, and it moves ahead of the subcluster as it propagates through the  main cluster. After crossing the pericenter, the subcluster is decelerated and eventually pulled backwards by the main cluster's gravity. The shock, however, continues to propagate all the way to larger cluster radii. In this \emph{detached} phase, the bow shock evolves into a ``runaway shock'', whose propagation is mainly determined by the gas density/pressure profiles of the main cluster. Eventually, this runaway merger shock overtakes the accretion shock and re-shapes the boundary of the ICM \citep{Zhang2020}.

In this study, we discuss additional ``orphan'' shocks that form in the central region of the main cluster long after the pericenter passage by the subcluster, and, therefore, the association of the shock with the subcluster is not obvious. In simulations, such shocks are often found a few $\Mpc$ behind the front of the runaway shock. In spite of such a large distance, these two shocks together constitute a giant, Mpc-scale, {\em N}-wave, which is named for its {\em N}-shape waveform  (\citealt{Dumond1946,Whitham1950,Whitham1974}; see Fig.~\ref{fig:1d_sph_waveform} for an example). The runaway merger shock and the orphan shock are the leading and trailing edges of the {\em N}-wave, respectively. Though {\em N}-waves have been extensively studied motivated by their importance for industrial applications (e.g. sonic booms driven by aircrafts; see \citealt{Plotkin1989}), their possible presence in galaxy clusters has not been discussed. In addition, gravitational potential/stratification of the gas in clusters modifies the properties of these waves. The aim of this paper is to draw attention to the formation and evolution of {\em N}-waves in the context of galaxy clusters.

This paper is organized as follows. In Section~\ref{sec:1d}, we discuss properties of idealized one-dimensional (1D) {\em N}-waves in a stratified atmosphere before proceeding with the analysis of a similar but more complicated process occurring in galaxy clusters. In Section~\ref{sec:3d}, we specifically explore the formation and evolution of {\em N}-waves in merging clusters and their dependence on merger configurations. In Section~\ref{sec:observation}, we present observational examples/candidates of {\em N}-waves in the Coma cluster and A2744. Finally, Section~\ref{sec:conclusions} summarizes our findings.

\section{Formation of 1D {\em N}-waves} \label{sec:1d}

In this section, we illustrate the formation of {\em N}-waves in stratified atmospheres in 1D simulations, which provide a useful analogue of the process occurring in merging clusters. In these simulations, we assume that the initial isothermal gas atmosphere is in hydrostatic equilibrium in a static gravitational potential. The initial gas density profile\footnote{Unless stated otherwise, we use the subscript $0$ in gas density $\rho_{\rm gas,0}$ (and also pressure $p_{\rm gas,0}$ and sound speed $c_{\rm s,0}$) to indicate the unperturbed state of the atmosphere. } follows
\be
\rho_{\rm gas,0}(r)=\left(\frac{r}{r_{\rm core}}+1\right)^{-\omega},
\label{eq:1d_sim_profile}
\ee
where $r_{\rm core}$ is the core radius\footnote{In 1D Cartesian coordinates (see Section~\ref{sec:1d:cartesian}), $r\equiv|x|$ in Eq.~(\ref{eq:1d_sim_profile}).}. The static gravitational potential is chosen to maintain hydrostatic equilibrium for the initial profile.
This profile is asymptotically a power-law when $r\gg r_{\rm core}$. For convenience, we set the initial gas pressure profile $p_{\rm gas,0}(r)=\rho_{\rm gas,0}(r)$ in our 1D simulations. Therefore, the sound speed of the atmosphere is $c_{\rm s,0}=\sqrt{\gamma}$, where $\gamma\ (=5/3)$ is the gas adiabatic index. With these definitions and for a cluster with a temperature $1\keV$, if one sets the code units of length to $100\kpc$, the code units of time would correspond to $243\Myr$.

In the standard picture of the {\em N}-wave formation, the non-linearity of the waves plays a central role in shaping the waveform. However, in our problem, the spherical geometry and the presence of the gravitational potential well both make non-negligible contributions to this process. We illustrate the impact of these effects on the formation of the {\em N}-wave in the following subsections.

\subsection{Effects of gravitational potential/stratification} \label{sec:1d:spherical}

We begin with the analysis of our spherically symmetric simulations, where we initiate a blast wave at the origin at $t=0$ by increasing pressure in the innermost cell ($r<10^{-2}$) by a factor of $\xi$. For each run, we select $\xi$ so that the Mach number of the leading shock front is $\mathcal{M}_{\rm s}\simeq1.4$ at $t=0.5$ (see the left panel in Fig.~\ref{fig:1d_sph_w}). Table~\ref{tab:1d_params} lists the parameters used in the simulations.

\begin{table}
\centering
\begin{minipage}{0.45\textwidth}
\centering
\caption{Parameters of 1D blast-wave simulations (see Section~\ref{sec:1d}).}
\label{tab:1d_params}
\begin{tabular}{ccccc}
  \hline
  IDs & Geometry  & $\omega$ & $r_{\rm core}$ & $\xi$  \\ \hline
   S0 & spherical & $0$ & $0.1$ & $3.0\times10^6$  \\
   S1 & spherical & $1$ & $0.1$ & $1.5\times10^5$  \\
   S2 & spherical & $2$ & $0.1$ & $1.0\times10^4$  \\
   S3 & spherical & $3$ & $0.1$ & $1.2\times10^3$  \\
   C2 & Cartesian & $2$ & $0.3$ & -- \\
\hline
\vspace{-3mm}
\end{tabular}
\end{minipage}
\end{table}

\begin{figure}
\centering
\includegraphics[width=0.9\linewidth]{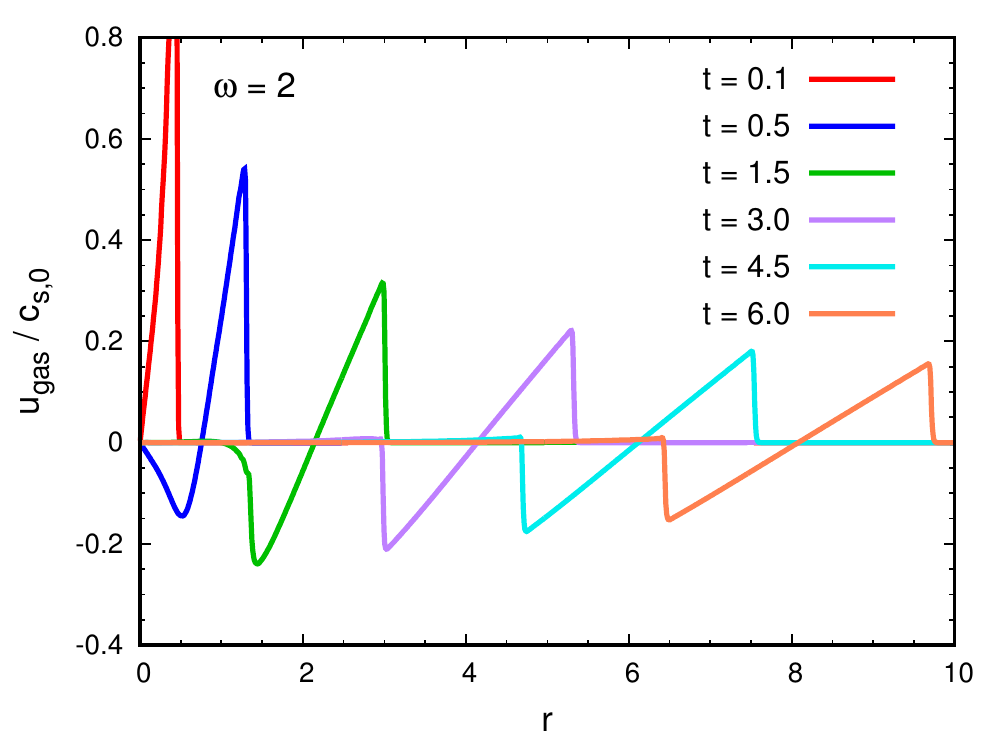}
\caption{Evolution of the gas velocity profiles (in units of the sound speed of unperturbed gas) in the spherical-1D simulation S2. This figure illustrates the formation of an {\em N}-wave in a stratified atmosphere with a power-law density profile (see Section~\ref{sec:1d:spherical}).}
\label{fig:1d_sph_waveform}
\end{figure}

\begin{figure}
\centering
\includegraphics[width=0.9\linewidth]{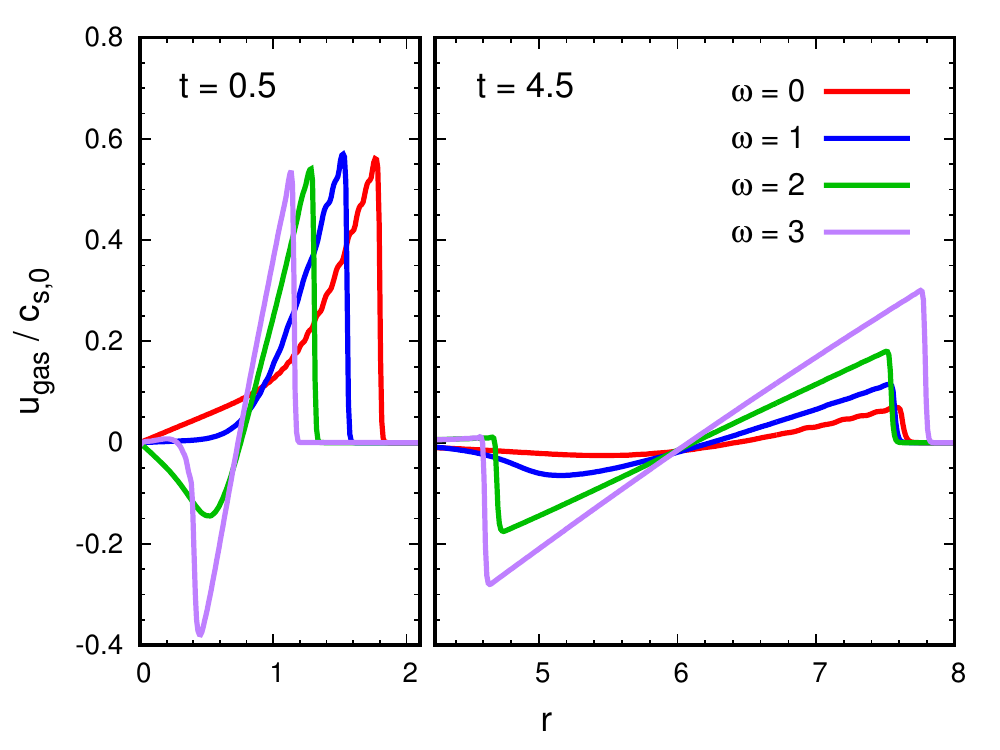}
\caption{Comparisons of the gas velocity profiles in the spherical-1D simulations with different $\omega$ at $t=0.5$ (left panel) and $t=4.5$ (right panel), respectively. This figure shows that gravity/stratification has a strong effect on the shape of the waves. The {\em N}-waves are formed on a shorter timescale if the atmosphere's density profile is steeper (see Section~\ref{sec:1d:spherical}).}
\label{fig:1d_sph_w}
\end{figure}

Fig.~\ref{fig:1d_sph_waveform} shows the formation of an {\em N}-wave from a blast wave in the simulation S2. The curves show the evolution of the gas velocity profiles $u_{\rm gas}$ in units of the sound speed of undisturbed gas $c_{\rm s,0}$. At the beginning of the simulation ($t\simeq0.5$), the outgoing blast wave is trailed by a rarefaction formed due to the spherical symmetry of the system. As the wave moves outwards, its leading shock front attenuates, while the rarefaction sharpens rapidly. A prominent {\em N}-wave is formed at $t\sim3$, which shows a nearly symmetric shape around the middle point where $u_{\rm gas}=0$. When the {\em N}-wave is far from the place of its origin, its behavior is close to the one in an asymptotic weak-shock solution in a stratified spherically-symmetric atmosphere with a power-law density profile, i.e. the wave's amplitude decreases as $\propto r^{\omega/4-1}$ while its characteristic length scale increases as $\propto r^{\omega/4}$ ($0<\omega\leq4$, see eq.~3 in \citealt{Zhang2019b}).

Fig.~\ref{fig:1d_sph_w} shows a comparison of the waveforms generated in simulations with different initial gas density profiles. In the uniform medium ($\omega=0$), the blast wave needs a very long time to evolve into an {\em N}-wave. Only the wave's non-linearity works here. In a stratified atmosphere, the timescale is significantly shortened. It implies that gravity plays an important role in shaping the waves. Generally, two main processes contribute to this effect. (1) The steep gas density profile helps to maintain the strength of the traveling shocks/waves in the atmosphere \citep{Zhang2019b}. In such a situation, the {\em N}-waves develop faster because the waves' non-linear timescale is inversely proportional to their amplitudes. (2) The gravitational potential/stratification strongly affects the development of waves in the central region, where the density/pressure scale height of the gas atmosphere is the smallest. The transition from the core region to a steeply declining density profile plays a particularly important role. The left panel in Fig.~\ref{fig:1d_sph_w} shows that the profile of the rarefaction is steeper and narrower when $\omega$ is larger.

One obvious limitation of the spherically symmetric model is its inner boundary condition, i.e. zero gas velocity at $r=0$, which is, of course, unrealistic for merging clusters. In the next subsection, we perform additional 1D simulations in Cartesian coordinates to avoid this issue and specifically explore the effect of the central density peak on the formation of the rarefaction.

\subsection{Core passage of waves} \label{sec:1d:cartesian}

\begin{figure}
\centering
\includegraphics[width=0.9\linewidth]{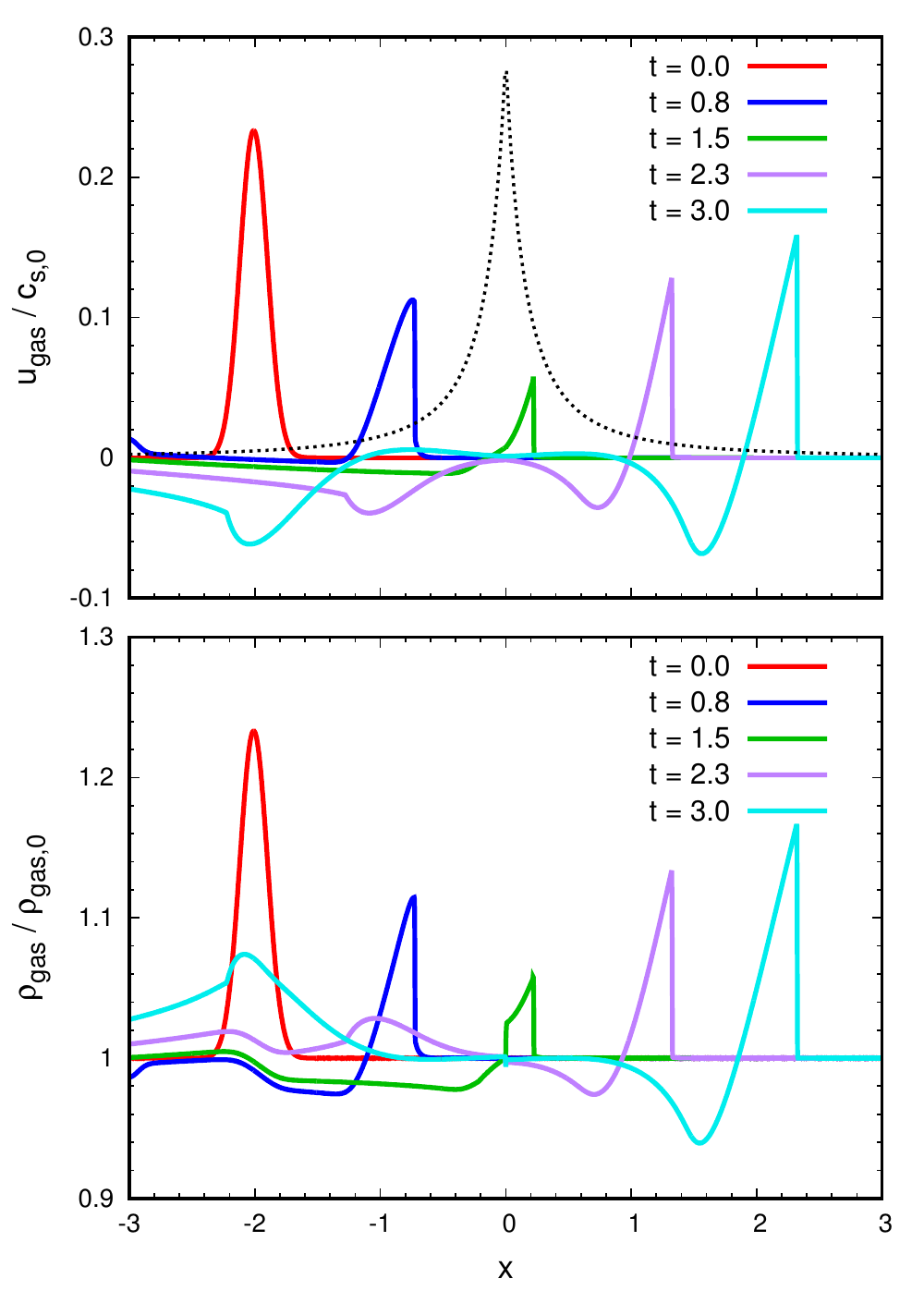}
\caption{Evolution of the gas velocity (top panel) and density (bottom panel) profiles of a linear wave in the Cartesian-1D simulation C2. The black dotted line in the top panel shows the shape of the unperturbed atmosphere density profile $\rho_{\rm gas,0}$. After crossing the origin ($x=0$), a steep and narrow rarefaction is formed behind the leading shock front, which is similar to that shown in Fig.~\ref{fig:1d_sph_waveform} (see Section~\ref{sec:1d:cartesian}). }
\label{fig:1d_cart}
\end{figure}

Instead of a spherical shock, we now consider a shock propagating through a stratified atmosphere in Cartesian-1D geometry.  We use the same form of the static gravitational potential as in Section~\ref{sec:1d:spherical} (see Eq.~\ref{eq:1d_sim_profile} and also Table~\ref{tab:1d_params} for the parameters used in the simulation), but set different initial perturbations. In the initial conditions, we generate a linear wave centered at $x=-2$, which has a shape of a narrow but smooth peak in the gas velocity profile (see the red line in the top panel of Fig.~\ref{fig:1d_cart}).
The gas density and pressure profiles of this wave are set to mimic an adiabatic sound wave, so that the density perturbation is proportional to the velocity perturbation, i.e. $(\rho_{\rm gas}-\rho_{\rm gas,0})/\rho_{\rm gas,0}=u_{\rm gas}/c_{\rm s,0}$ and $p_{\rm gas}/p_{\rm gas,0}=\left (\rho_{\rm gas}/\rho_{\rm gas,0}\right)^\gamma$ while the entropy of gas equals to its value in the unperturbed profile.

\begin{figure}
\centering
\includegraphics[width=0.9\linewidth]{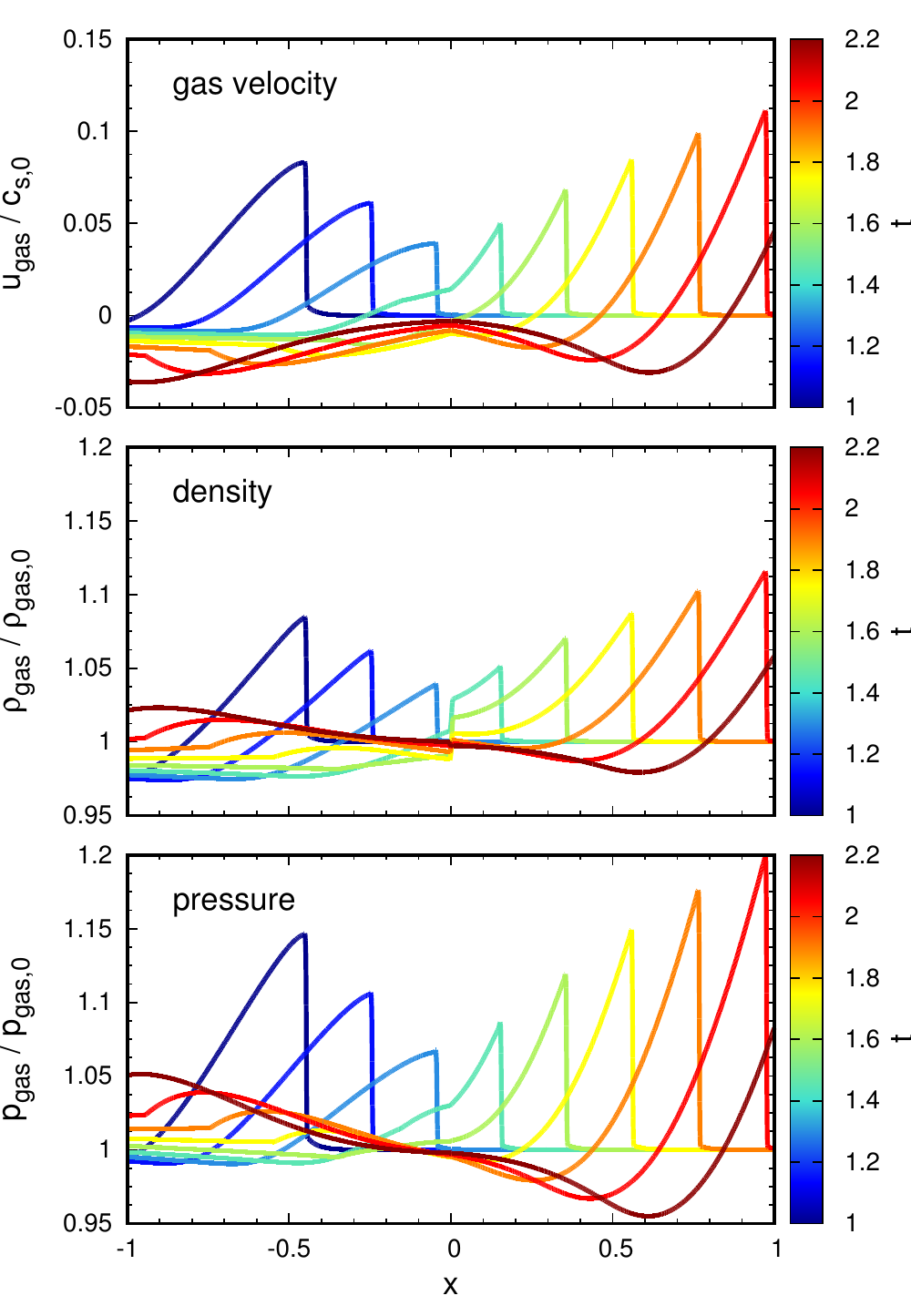}
\caption{Evolution of the gas velocity, density, and pressure perturbations in the core of the atmosphere in the Cartesian-1D simulation C2 during the passage of the shock. Near the origin, the gas quickly ($\triangle t\lesssim1$) restores its initial state. After passing the core, a rarefaction develops, which trails the shock front as it moves rightwards (see Section~\ref{sec:1d:cartesian}).}
\label{fig:1d_cart_zoom}
\end{figure}

Fig.~\ref{fig:1d_cart} shows the evolution of the gas velocity and density profiles in the simulation C2. As the wave approaches the center, its amplitude goes down, largely due to density increase towards $x=0$. The wave is then partially reflected while crossing the core of the atmosphere at $x=0$. The reflected part associated with a negative gas velocity moves leftwards. The transmitted part develops a prominent rarefaction behind the leading front of the wave after $t\simeq2$ (see also Fig.~\ref{fig:1d_cart_zoom} for a zoomed view on the central region), which is similar to what we have seen in Fig.~\ref{fig:1d_sph_waveform}.

Fig.~\ref{fig:1d_cart_zoom} shows the evolution of the gas properties (i.e. velocity, density, and pressure) zooming in on the central region of the atmosphere for a clearer view of the waves' behavior at $x\simeq0$. After the shock front crosses the core, a steep edge (but not a discontinuity) appears in the density profile near $x=0$. The shock slightly shifts the gas atmosphere rightwards relative to the static potential well. The minimum of the entropy profile is now located at the right end of this edge ($x>0$) rather than at $x=0$. We can also see that the perturbed atmosphere restores quickly to its initial state near the center of the gravitational potential. Since the acoustic cut-off frequency of the atmosphere is a decreasing function of the radius, i.e. $\omega_{\rm ac}\propto1/(r+r_{\rm core})$, the long-wavelength components (i.e. larger than the pressure scale height) of the perturbations have smaller group velocities and tend to be confined in the central region of the atmosphere. Their typical restoring timescale is close to $T_{\rm ac}=2\pi/\omega_{\rm ac}$, approximately $T_{\rm ac}\simeq1.5$ at $x=0$ in the run C2. On the contrary, the short-wavelength components of the perturbations propagate to the larger radii with the velocity $\sim c_{\rm s}$ and constitute the steep structures we have seen in the gas profiles \citep{Lamb1909,Kalkofen1994}. This effect explains why the presence of the gravitational potential/stratification promotes the formation of narrow and steep trailing rarefactions in our simulations.

In general, this simulation resembles 2D simulations done in \citet[][see their fig.~11 and section~4]{Churazov2003}, where a passage of a shock through the core creates a complicated pattern of gas motion, that involves both gas sloshing and shocks.

\section{\textit{N}-waves in merging clusters} \label{sec:3d}

Though our 1D models consistently lead to a formation of {\em N}-waves in stratified atmospheres, the process of clusters merger is much more complicated. To take some of these complexities into account (e.g. non-spherical geometry, variations of gravitational potential), we further performed 3D hydrodynamic simulations of idealized cluster mergers.

In these simulations of the mergers of two clusters, the initial conditions are very similar to those in \citet{Zhang2014,Zhang2015}. Each cluster is composed of spherical DM and gas halos that are both in equilibrium in the initial conditions (see section~2.1 in \citealt{Zhang2014} for more information on the simulation setup). For simplicity, we fixed the concentration parameter $c_{\rm vir}=4$ used in the DM density profile and the gas fraction within the virial radius $f_{\rm gas}=0.12$ for all our clusters.
The merger configurations in the simulations are determined by the masses of two merging clusters $M_1$ and $M_2$, initial relative velocity $V$, and impact parameter $P$. Table~\ref{tab:merger_params} summarizes these parameters used in our work. All simulations presented in this section are performed with the moving-mesh code {\sc Arepo} \citep{Springel2010,Weinberger2020}. The mass resolution of the DM and gas reaches $10^9$ and $10^8\msun$, respectively.
In Section~\ref{sec:3d:formation}, we analyze the process of the {\em N}-wave formation in the run M10M1P2, and further discuss the dependence of this process on cluster merger configurations in Section~\ref{sec:3d:parameters}.

\begin{table}
\centering
\begin{minipage}{0.45\textwidth}
\centering
\caption{Parameters of cluster-merger simulations (see Section~\ref{sec:3d}).}
\label{tab:merger_params}
\begin{tabular}{cccccc}
  \hline
  IDs & $M_1,\,M_2\,(10^{14}M_\odot)$ & $V\,({\rm km\,s^{-1}})$ & $P\,({\rm Mpc})$  \\\hline
  M10M1P0 & $10,\ 1$ & 500 & $0$   \\
  M10M1P2 & $10,\ 1$ & 500 & $2$ \\
  M10M1P4 & $10,\ 1$ & 500 & $4$ \\
  M6M3P0 & $6,\ 3$ & 500 & $0$ \\
  M6M3P2 & $6,\ 3$ & 500 & $2$ \\

%
\hline
\vspace{-3mm}
\end{tabular}
\end{minipage}
\end{table}

\subsection{Formation of {\em N}-waves in merging galaxy clusters} \label{sec:3d:formation}

\begin{figure*}
\centering
\includegraphics[width=0.95\linewidth]{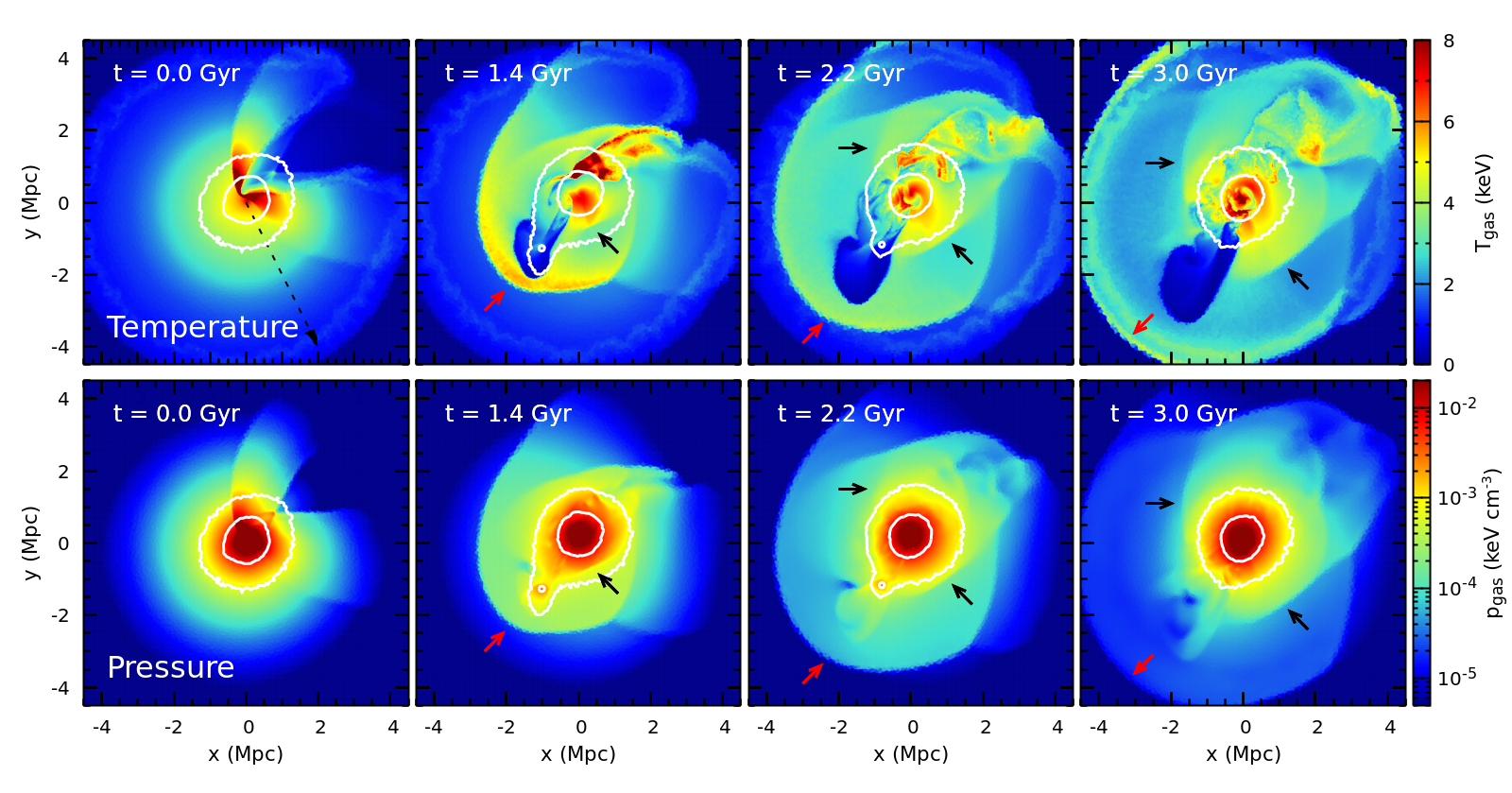}
\vspace{-3mm}
\caption{Gas temperature (top panels) and pressure (bottom panels) slices of a simulated merging cluster in the merger plane from the simulation M10M1P2. Total mass surface density is overlaid as white contours. The first two columns show the snapshots at the moments of the primary pericentric and apocentric passages of the infalling subcluster, respectively. The red arrows indicate the leading edge of the {\em N}-wave formed during the merger process (i.e. the runaway merger shock); and the black arrows indicate the trailing edge of the {\em N}-wave. This figure illustrates the formation of an {\em N}-wave in a merging cluster (see Section~\ref{sec:3d:formation}).}
\label{fig:merger_slice}
\end{figure*}

\begin{figure}
\centering
\includegraphics[width=0.9\linewidth]{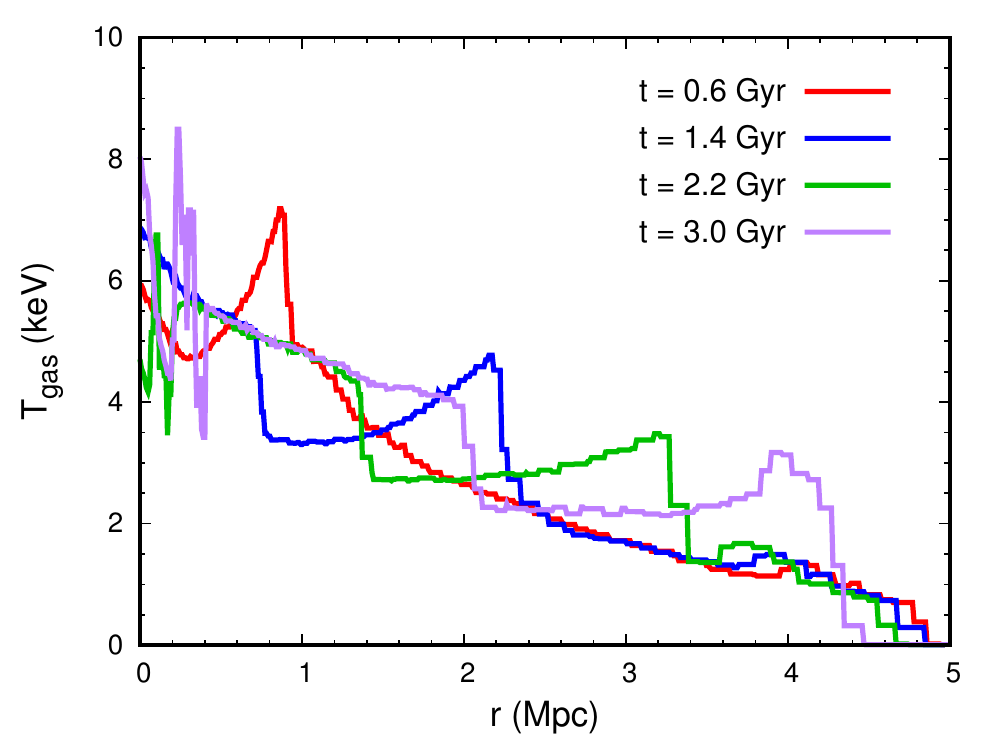}
\caption{Evolution of the gas temperature profiles in the simulation M10M1P2 along the direction indicated by the dashed arrow shown in Fig.~\ref{fig:merger_slice} (top-left panel). An {\em N}-wave is formed at $t\simeq1.4\Gyr$ after a merger shock crossing the center of the main cluster. It shows a very similar picture to that seen in the 1D model (cf. Fig.~\ref{fig:1d_sph_waveform}; see Section~\ref{sec:3d:formation}).}
\label{fig:merger_prof}
\end{figure}

Fig.~\ref{fig:merger_slice} shows the evolution of the gas temperature and pressure slices in the merger plane ($x-y$ plane) in the simulation M10M1P2. The overlaid white contours show the distribution of the total mass surface density. We set $t=0$ at the moment of the primary pericentric passage (shown in the first column). For $t>0$, the merger shock detaches from the subcluster that initially drives it. The snapshot of the primary apocentric passage is shown in the second column, when the {\em N}-wave has already been formed, though its rear edge is not yet very sharp (see also Fig.~\ref{fig:merger_prof}). The leading and trailing edges of the {\em N}-wave are indicated by the red and black arrows, respectively. We note here that the {\em N}-wave is formed once the merger shock sweeps through the center of the main cluster. This is similar to the 1D simulations (see Section~\ref{sec:1d}), however, the {\em N}-wave here reveals an approximately semispherical geometry. The third and fourth columns show the growth of the {\em N}-wave, whose rear edge becomes clearer and the characteristic distance between the two shocks increases to $\sim2\Mpc$. In this simulation, the {\em N}-wave is able to survive for a very long time because the gas density profile of the atmosphere is as steep as $\sim r^{-3}$ in the cluster outskirts \citep{Zhang2019b}.

The cold gas blob trailing the re-infalling subcluster is an interesting feature seen in the top panels in Fig.~\ref{fig:merger_slice}. It is the gas stripped from the subcluster while crossing the main cluster. This stripped gas experiences adiabatic expansion while radially moving outwards since the atmosphere has a large pressure gradient in the cluster outer region. Eventually, most of this cold and low-entropy gas falls back towards the center of the main cluster. This cold-gas flow, however, interferes with the {\em N}-wave. Therefore, it is difficult to see in these slices the trailing edge of the {\em N}-wave along the directions connecting the subcluster and the main cluster center.

Fig.~\ref{fig:merger_prof} shows the evolution of the temperature profiles along the direction indicated by the black dashed arrow in Fig.~\ref{fig:merger_slice} (top-left panel), which gives a clear view of the {\em N}-wave formation in merging clusters. At $t=0.6\Gyr$, the runaway merger shock is followed by a rarefaction, which quickly evolves into a shock front within $\sim1\Gyr$. This behavior is similar to that seen in the spherical-1D model (cf. Fig.~\ref{fig:1d_sph_waveform})\footnote{Note, in our spherical-1D models, an {\em N}-wave shows similar waveforms in its gas temperature and velocity profiles.}. The {\em N}-wave is formed at $t\simeq1.4\Gyr$ and propagates radially outwards. In contrast with the 1D model, the leading shock front of the {\em N}-wave here is slightly stronger than the trailing one.

\subsection{Dependence on merger configurations} \label{sec:3d:parameters}

A natural question is how common are the {\em N}-waves in galaxy clusters? To gain insights into this question, we investigated the dependence of {\em N}-wave formation on the cluster merger configurations by varying merger parameters in the simulations. For simplicity, we fix the initial relative velocity between the two merging clusters (i.e. $V=500\kms$, a typical cluster pairwise velocity motivated by cosmological simulations; see e.g. \citealt{Thompson2012,Dolag2013}) but vary the cluster masses and impact parameter in the runs.

\begin{figure*}
\centering
\includegraphics[width=0.95\linewidth]{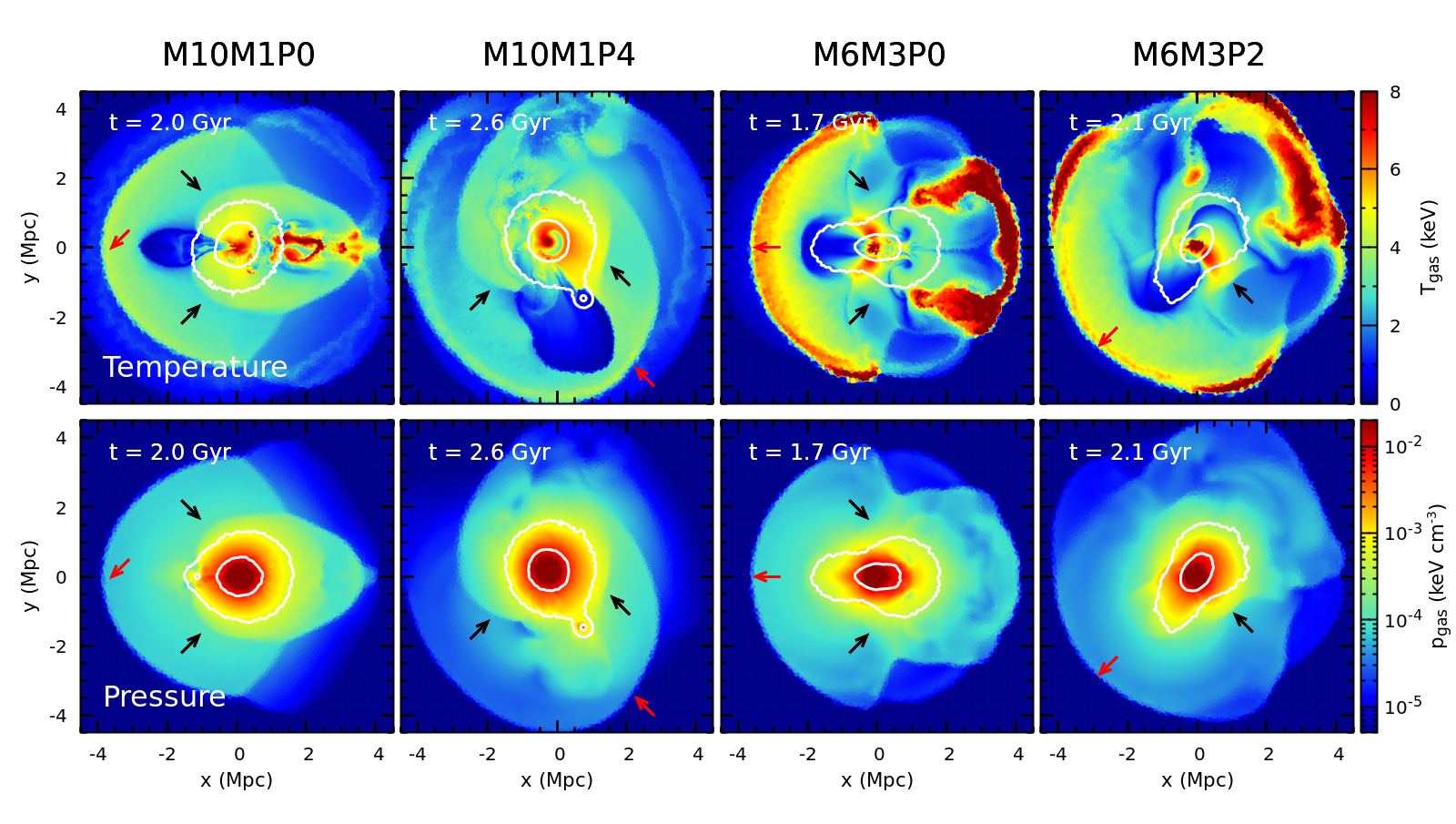}
\vspace{-3mm}
\caption{Similar to Fig.~\ref{fig:merger_slice} but the snapshots from the simulations with different merger configurations. This figure shows that {\em N}-waves are very commonly formed in merging clusters, though they are more prominent in the clusters undergoing minor mergers (see Section~\ref{sec:3d:parameters}). }
\label{fig:params_explore}
\end{figure*}

Fig.~\ref{fig:params_explore} shows selected snapshots of these simulations at the moments within $1\Gyr$ since the primary apocentric passage. The panels show the temperature (top) and pressure (bottom) slices in the merger plane. We can see that {\em N}-waves appear in all four runs, marked by the red and black arrows as in Fig.~\ref{fig:merger_slice}. However, they are more significant in clusters undergoing minor mergers with the merger mass ratio $M_1/M_2=10$. The impact parameter, on the contrary, shows little effect on {\em N}-waves, though the waves have a more symmetric shape in the head-on mergers ($P=0$). In major mergers ($M_1/M_2=2$), a large fraction of the ICM is re-distributed by the violent gas motions, which makes the {\em N}-waves less prominent on top of other complicated gaseous structures.

\section{Observational candidates} \label{sec:observation}

In the previous sections, we have demonstrated that pairs of shocks (giant {\em N}-waves) naturally form in merging clusters. It is, therefore, interesting to search for such shocks in observed galaxy clusters. Guided by our simulations, we look for these waves in clusters that are undergoing minor mergers ($M_1/M_2\sim10$) and have an infalling subcluster near the apocenter.

\begin{figure*}
\centering
\includegraphics[width=\linewidth]{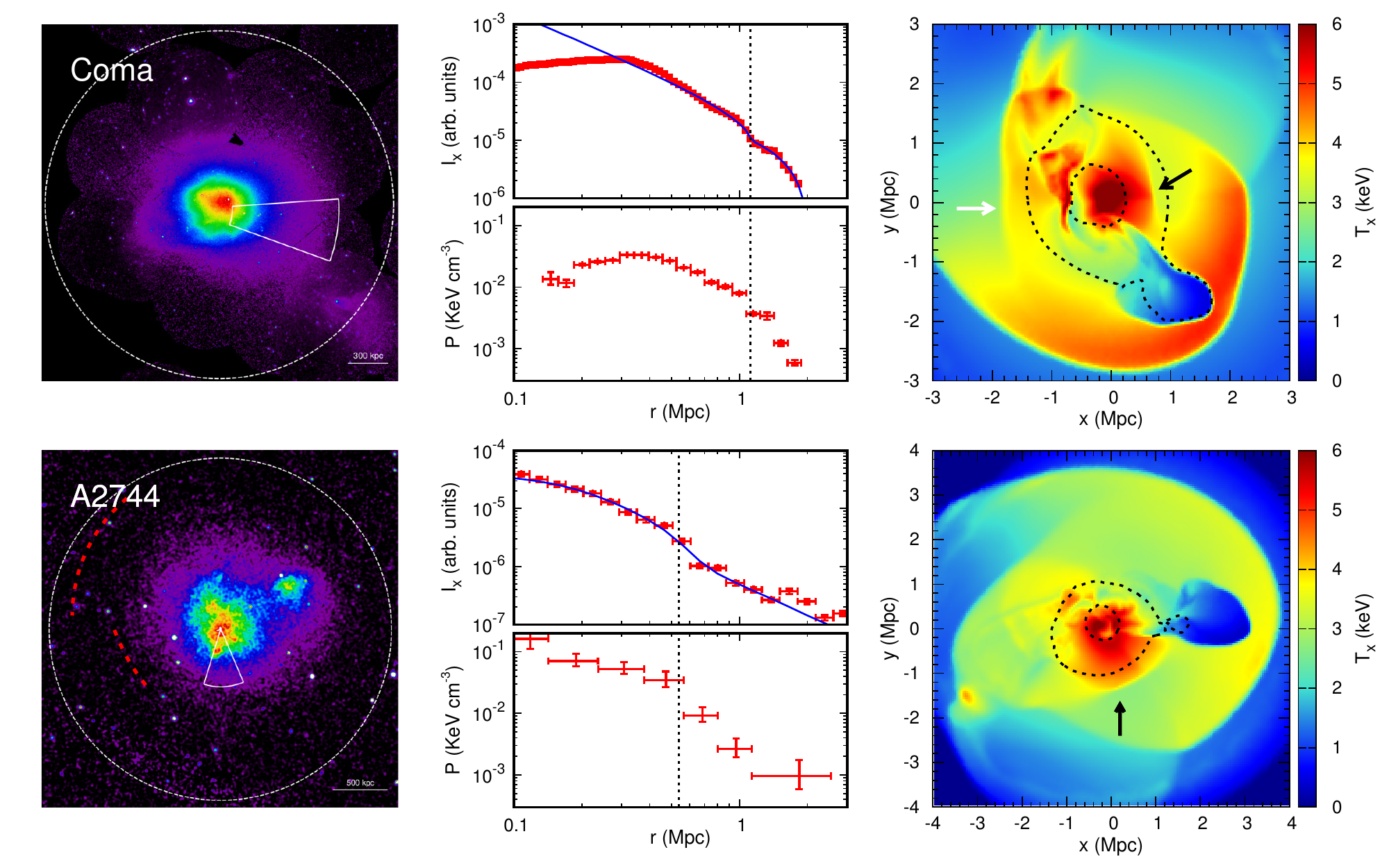}
\vspace{-3mm}
\caption{{\em N}-wave candidates identified in the Coma cluster (top panels) and A2744 (bottom panels). The left panels show the X-ray SB images. The white dashed circles indicate the cluster $R_{500}$. The red dashed curves in the bottom-left panel approximately indicate the positions of the radio relics shown in \citet{Pearce2017}. The SB and thermal pressure profiles extracted in the white sectors are shown in the middle panels, respectively. The SB profiles are further fitted by a broken power-law density model (blue lines). The best-fit shock radii are indicated by the vertical dashed lines. As a comparison, the right panels show the X-ray-weighted temperature distributions (see Eq.~\ref{eq:xray_temp}) estimated in the simulation M10M1P2 at $t=1.4$ and $2.1\Gyr$, respectively. They illustrate possible merger configurations of Coma and A2744 (see Section~\ref{sec:observation}). The overlaid black dashed contours show the distribution of the modeled X-ray SB. The trailing edges of the {\em N}-waves are indicated by the black arrows (see also Fig.~\ref{fig:merger_slice}). The white arrow in the top-right panel marks the runaway shock driven by the main cluster, which may correspond to the southeast shock identified in \citet[][see their fig.~2]{Planck2013}. }
\label{fig:xray_obs}
\end{figure*}

Two promising candidates, the Coma cluster and A2744, are shown in Fig.~\ref{fig:xray_obs}. For the Coma cluster, we used publicly available \textit{XMM-Newton} data from the EPIC/MOS detector. The data were cleaned following the procedure described in \citet{Churazov2003}. For A2744, we utilized archived \textit{Chandra}/ACIS data. The data were prepared using the algorithms described in \citet{Vikhlinin2005}. The left panels in Fig.~\ref{fig:xray_obs} show the background-subtracted, exposure- and vignetting-corrected mosaic images of both clusters. Excluding the point sources from the images, we extracted X-ray surface brightness (SB) profiles in the sector $340^\circ - 364^\circ$ ($253^\circ - 293^\circ$) centered at RA: 195.11 (3.57) and DEC: 27.91 (-30.39) in Coma (A2744).
Both SB profiles are then fitted with the broken power-law model to obtain the best-fitting position of the SB jumps (dashed lines in the middle panels in Fig.~\ref{fig:xray_obs}). We deprojected spectra using the algorithm described in \citet{Churazov2003} and fitted the spectra to obtain gas density and temperature using XSPEC and a single-temperature model \citep{Smith2001,Foster2012} based on ATOMDB version 3.0.9. We obtained the thermal pressure as a product of projected temperature and the total number density of particles. Fig.~\ref{fig:xray_obs} shows that jumps in the SB profiles correspond to the jumps in pressure, which guarantees that the detected edges are shocks rather than contact discontinuities. These shocks are consistent with previous findings \citep{Planck2013,Owers2011}.

To qualitatively illustrate the merger configurations of these two clusters, we further show the X-ray-weighted temperature distributions estimated in our simulation M10M1P2 in the right panels in Fig.~\ref{fig:xray_obs}, i.e.
\be
T_{\rm x} = \frac{\int_{\rm LOS}{T_{\rm gas}\varepsilon_{\rm x}\dd\ell}}{\int_{\rm LOS}{\varepsilon_{\rm x}\dd\ell}},
\label{eq:xray_temp}
\ee
where $\varepsilon_{\rm x}$ is the X-ray bolometric emissivity (see eq.~7 in \citealt{Zhang2014}). For comparison with the observations, we rotate the maps along the $z-$axis and/or flip them vertically. The line-of-sight (LOS), however, is always perpendicular to the merger plane, even though such a condition is not necessary to reveal the {\em N}-waves in the map. We emphasize that we did not attempt to fine-tune the merger parameters and viewing angles and quantitatively ``fit'' the observations, but rather to illustrate plausible merger scenarios for Coma and A2744 in terms of morphology of the X-ray SB, the relative positions of the main and sub clusters and the shock fronts.

In the Coma cluster, the infalling group (i.e. NGC~4839) is close to its apocenter \citep{Lyskova2019,Sheardown2019}. The leading edge of the runaway shock coincides with the radio relics discovered in the cluster southwest \citep{Brown2011}. Our numerical model shows that the trailing edge of the {\em N}-wave is on the west of the cluster (marked by the black arrow in the right panel of Fig.~\ref{fig:xray_obs}). In observations, such a shock front is detected along this direction as the model predicts in both the X-ray image (see the top-middle panel) and the SZ signals \citep{Planck2013}. Our {\em N}-wave scenario suggests that this west shock may have the same origin as that of the radio relics, though their projected distance is up to $\sim1\Mpc$. Interestingly, \citet{Planck2013} has found another shock front in the southeast of the cluster. This shock corresponds to the leading edge of the runaway shock driven by the main cluster in our model (marked by the white arrow). Its morphology, however, is partially affected by the wake of the subcluster.

A2744 exhibits a complex merger configuration \citep[e.g.][]{Owers2011,Merten2011,Medezinski2016} and also rich extended radio sources \citep[see][and references therein]{Pearce2017}. We here focus on its northwestern substructure seen in the X-ray image. We explain it as due to a small subcluster that has crossed its apocenter and is now falling back to the center of the main cluster \citep[see also][]{Kempner2004,Merten2011}. In our model, the position of the runaway merger shock, driven by the main cluster in this minor merger process, coincides with the discovered radio relics to the east from the cluster (see the bottom panels in Fig.~\ref{fig:xray_obs} and also \citealt{Pearce2017}). The trailing edge of the {\em N}-wave is expected to be to the south from the main cluster (marked by the black arrow). In observations, a shock does exist along this direction (see the bottom-middle panel in Fig.~\ref{fig:xray_obs}) and may correspond to the one predicted in our simulation. However, we note that a bow shock associated with the ``southern compact core'' in A2744 may also be located in this region \citep[][see their fig.~1]{Owers2011}. Since the {\em N}-wave is formed earlier, it is possible that this bow shock has already overtaken the trailing edge of the {\em N}-wave.

Besides those exampled above, there are other interesting candidates for the search of {\em N}-wave structures, for example, A85 \citep[e.g.][]{Ichinohe2015} and A2142 \citep[e.g.][]{Markevitch2000,Eckert2014}. Their studies may further guide the search of the corresponding leading edges of the runaway shocks in cluster peripheries and the relevant radio signals.

\section{Conclusions} \label{sec:conclusions}

In this study, we numerically explored the formation of Mpc-scale {\em N}-waves in merging clusters and discussed two observational candidates in the Coma cluster and A2744.

Through numerical simulations, we showed that a merger shock gradually detaches from the subcluster that initially drives it, after the core passage, and moves all the way to the cluster outskirts \citep[see also][]{Zhang2019b}. This runaway merger shock is trailed by a rarefaction, which quickly evolves into a second shock front due to the non-linearity. This secondary shock is much closer to the main cluster center and constitutes a giant {\em N}-shaped wave with the leading front of the runaway shock. Despite being a close analogue of the classical {\em N}-waves known in hydrodynamics, this structure is formed in a much shorter timescale in the cluster environment due to the impact of the atmosphere stratification in the cluster gravitational potential. The distance between the leading and trailing shocks could be up to a few Mpc.

Since the leading fronts of the fully developed {\em N}-waves (i.e. runaway merger shocks) are always at large cluster radii, it might be challenging to detect them in existing X-ray observations. However, detecting their trailing shocks should be much easier. Two candidates potentially possessing {\em N}-waves -- the Coma and A2744 clusters are discussed in this context (see Section~\ref{sec:observation}). These trailing-front candidates potentially provide observational clues for the presence of their associated runaway merger shocks lying in the periphery of the clusters. The latter might be bright in the synchrotron emission in radio band, like those seen in the Coma cluster.

By numerically exploring different cluster merger configurations, we showed that the {\em N}-waves are expected to be ubiquitous in galaxy clusters. They are particularly prominent in minor mergers with the mass ratio $\simeq10$. In this sense, it is interesting to systematically search for such structures in well-studied merger clusters, the work we defer for future publications.

\section*{Acknowledgments}

Part of the simulations presented in this paper were carried out using the Midway computing cluster provided by the University of Chicago Research Computing Center. IZ is partially supported by a Clare Boothe Luce Professorship from the Henry Luce Foundation. EC acknowledges support by the Russian Science Foundation grant~19-12-00369.

\section*{Data Availability}

The data underlying this article will be shared on reasonable request to the corresponding author.

\bsp	
\label{lastpage}
\end{document}